\documentclass[aip,amsmath,amssymb,reprint]{revtex4-2}
\usepackage{graphicx}
\usepackage{dcolumn}
\usepackage{bm}
\usepackage[utf8]{inputenc}
\usepackage[T1]{fontenc}
\usepackage{mathptmx}
\usepackage{etoolbox}
\usepackage{soul,xcolor}
\makeatletter
\def\@email#1#2{%
 \endgroup
 \patchcmd{\titleblock@produce}
  {\frontmatter@RRAPformat}
  {\frontmatter@RRAPformat{\produce@RRAP{*#1\href{mailto:#2}{#2}}}\frontmatter@RRAPformat}
  {}{}
}
\makeatother

\begin{document}


\title[]{Broadband Bandpass Purcell Filter for Circuit Quantum Electrodynamics}
\author{Haoxiong Yan}
\affiliation{Pritzker School of Molecular Engineering, University of Chicago, Chicago IL 60637, USA}

\author{Xuntao Wu}
\affiliation{Pritzker School of Molecular Engineering, University of Chicago, Chicago IL 60637, USA}

\author{Andrew Lingenfelter}
\affiliation{Pritzker School of Molecular Engineering, University of Chicago, Chicago IL 60637, USA}
\affiliation{Department of Physics, University of Chicago, Chicago IL 60637, USA}

\author{Yash J. Joshi}
\affiliation{Pritzker School of Molecular Engineering, University of Chicago, Chicago IL 60637, USA}

\author{Gustav Andersson}
\affiliation{Pritzker School of Molecular Engineering, University of Chicago, Chicago IL 60637, USA}

\author{Christopher R. Conner}
\affiliation{Pritzker School of Molecular Engineering, University of Chicago, Chicago IL 60637, USA}

\author{Ming-Han Chou}
\affiliation{Pritzker School of Molecular Engineering, University of Chicago, Chicago IL 60637, USA}
\affiliation{Department of Physics, University of Chicago, Chicago IL 60637, USA}

\author{Joel Grebel}
\affiliation{Pritzker School of Molecular Engineering, University of Chicago, Chicago IL 60637, USA}

\author{Jacob M. Miller}
\affiliation{Pritzker School of Molecular Engineering, University of Chicago, Chicago IL 60637, USA}
\affiliation{Department of Physics, University of Chicago, Chicago IL 60637, USA}

\author{Rhys G. Povey}
\affiliation{Pritzker School of Molecular Engineering, University of Chicago, Chicago IL 60637, USA}
\affiliation{Department of Physics, University of Chicago, Chicago IL 60637, USA}

\author{Hong Qiao}
\affiliation{Pritzker School of Molecular Engineering, University of Chicago, Chicago IL 60637, USA}

\author{Aashish A. Clerk}
\affiliation{Pritzker School of Molecular Engineering, University of Chicago, Chicago IL 60637, USA}

\author{Andrew N. Cleland}
\affiliation{Pritzker School of Molecular Engineering, University of Chicago, Chicago IL 60637, USA}
\affiliation{Argonne National Laboratory, Lemont IL 60439, USA}
\date{\today}

\begin{abstract}
In circuit quantum electrodynamics (QED), qubits are typically measured using dispersively-coupled readout resonators. Coupling between each readout resonator and its electrical environment however reduces the qubit lifetime via the Purcell effect. Inserting a Purcell filter counters this effect while maintaining high readout fidelity, but reduces measurement bandwidth and thus limits multiplexing readout capacity. In this letter, we develop and implement a multi-stage bandpass Purcell filter that yields better qubit protection while simultaneously increasing measurement bandwidth and multiplexed capacity. We report on the experimental performance of our transmission-line--based implementation of this approach, a flexible design that can easily be integrated with current scaled-up, long coherence time superconducting quantum processors.
\end{abstract}

\maketitle

Circuit quantum electrodynamics (QED) provides a scalable approach to quantum information processing, using Josephson-based circuits as qubits \cite{Blais2021}. A popular way to measure the quantum state of a superconducting qubit is to probe the state-dependent frequency shift of a readout resonator dispersively coupled to the qubit \cite{Wallraff2005,Koch2007}. To realize fast readout, the resonator-environment coupling must be large, so that the resonator can rapidly absorb and emit measurement photons. However, the qubit relaxation time $T_1$ is then limited by the resulting Purcell effect \cite{Purcell1995}. To overcome this, Purcell filters have been developed, which suppress qubit emission by engineering the electrical environment seen by the readout resonator \cite{Reed2010,Jeffrey2014,Sank2014,Sete2015,Walter2017,Sunada2022,Chen2023}, using for example a single-pole bandpass filter \cite{Jeffrey2014}. This however limits the measurement bandwidth and thus the number of readout resonators that can be measured via a single readout line, and further is incompatible with long qubit relaxation times \cite{Place2021,Wang2022}. For multiplexed qubit readout, one solution is to design a separate Purcell filter for each readout resonator \cite{Heinsoo2018,Saxberg2022}, but this requires good frequency matching between the filter and readout resonator, making design and fabrication more complex. A second solution is to increase the number of bandpass filter stages, yielding a broader passband and better isolation in the filter stopband \cite{Matthaei1980,Pozar2011}, an approach demonstrated using e.g. coupled $\lambda/4$ resonators \cite{Mohebbi2014}, stepped-impedance transmission lines \cite{Bronn2015,Li2023a}, coupled mechanical resonators \cite{Cleland2019}, and coupled $LC$ resonators \cite{Zhang2023,Ferreira2022}. These multi-stage designs are all symmetric, with equally-coupled input and output ports. Here we present designs for both symmetric and asymmetric bandpass filters appropriate for circuit QED, in the asymmetric case implementing different coupling rates for the input versus output ports \cite{Jeffrey2014}. Using the coupled-mode picture \cite{Matthaei1980,Naaman2022}, we show that the readout resonator coupling point to the filter must be chosen carefully, to accommodate interference between the filter stages. We demonstrate that asymmetric filters provide better qubit protection, and that broader passbands with better protection are achieved by adding filter stages. We then experimentally implement this approach and test a simple, robust design that only uses sections of transmission lines.

\begin{figure}
  \includegraphics[]{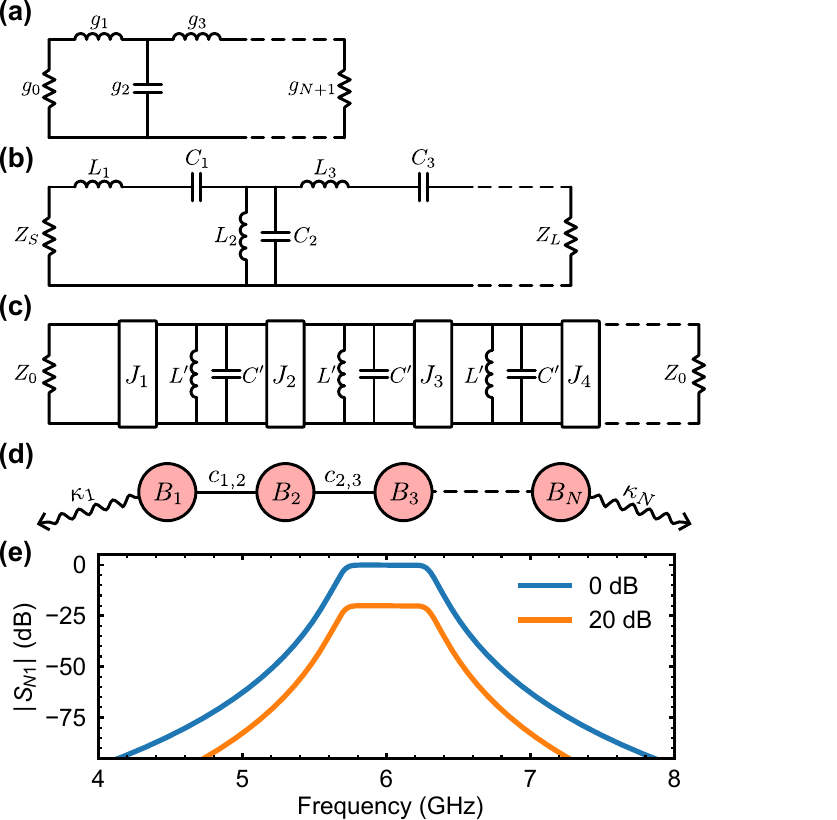}
  \caption{\label{fig1} Design flow for a bandpass filter. (a) Low-pass prototype, using capacitors and inductors. (b) Bandform transformation from lowpass to bandpass. (c) Bandpass filter with admittance inverters $J_n$, using only parallel $LC$ resonators. (d) Coupled-mode picture for $N$th order filter, with dissipation rate $\kappa_j$ for the $j$th resonator $B_j$ and coupling $c_{j, j+1}$ between resonators $B_j$ and $B_{j+1}$. (e) Transmission coefficient $|S_{N1}|$ for 6th-order bandpass filter designs with $0~\mathrm{dB}$ (blue) and $20~\mathrm{dB}$ (orange) insertion loss, center frequency $\omega_0/2\pi=6~\mathrm{GHz}$ and bandwidth $\Delta \omega/2\pi =600~\mathrm{MHz}$.}
\end{figure}

\begin{table}
  \caption{\label{tab:table1}Design coefficients $g_j$ for maximally-flat low-pass filter prototypes, with $g_0=1$.}
  \begin{ruledtabular}
  \begin{tabular}{ccccccccc}
  Insertion loss&Order&$g_0g_1$&$g_1g_2$&$g_2g_3$&$g_3g_4$&$g_4g_5$&$g_5g_6$&$g_6g_7$\\
  \hline
  0 dB& 1 &2.000&2.000&&&&&\\
   & 2 &1.414&2.000&1.414&&&&\\
   & 3 &1.000&2.000&2.000&1.000&&&\\
   & 4 &0.765&1.414&3.414&1.414&0.765&&\\
   & 5 &0.618&1.000&3.236&3.236&1.000&0.618&\\
   & 6 &0.518&0.732&2.732&3.732&2.732&0.732&0.518\\
   \hline
   20 dB& 1 &399.0&1.003&&&&&\\
   & 2 &563.6&1.003&0.708&&&&\\
   & 3 &597.5&2.003&0.668&0.500&&&\\
   & 4 &607.6&2.417&1.709&0.415&0.383&&\\
   & 5 &615.2&2.621&2.344&1.237&0.277&0.309&\\
   & 6 &618.3&2.734&2.734&1.868&0.911&0.196&0.259\\
  \end{tabular}
  \end{ruledtabular}
\end{table}

We first illustrate our design flow for the bandpass filters (Fig.~\ref{fig1}). We begin with a low-pass filter prototype \cite{Matthaei1980,Pozar2011}, with the circuit for a normalized $N$th order low-pass filter shown in Fig.~\ref{fig1}(a), where $g_0$ ($g_{N+1}$) is the source (load) impedance number, with the source impedance normalized to $1~\mathrm{\Omega}$ and the filter bandwidth normalized to $\omega_c=1~\mathrm{rad/s}$. For a given filter response (maximally flat, equal ripple, etc.), the $g_j$ coefficients can be calculated using the insertion loss technique \cite{Pozar2011} and Cauer synthesis \cite{Matthaei1980,Naaman2022}. We list the $g_j$ coefficients for maximally flat, low-pass filter designs in Table~\ref{tab:table1}, for both conventional symmetric filters with $0~\mathrm{dB}$ insertion loss as well as for asymmetric filters with large insertion loss ($20~\mathrm{dB}$). There are no lossy elements in the filter circuit, thus the insertion loss comes from the reflection at the input port. A bandpass filter with large insertion loss corresponds to a filter with a weakly-coupled input port and a strongly-coupled output port, so that most of the driving signal reflects from the input port while most of the signal scattered from the readout resonator is emitted from the output port, improving the readout efficiency \cite{Jeffrey2014,Wang2021}.

Based on the low-pass prototype in Fig.~\ref{fig1}(a), we can obtain the bandpass filter circuit, with center frequency $\omega_0$ and bandwidth $\Delta \omega$, through frequency and impedance scaling and bandform transformations \cite{Matthaei1980,Pozar2011}, as shown in Fig.~\ref{fig1}(b). The prototype circuit contains both series and parallel $LC$ resonators, which can be difficult to realize at microwave frequencies. We therefore transform this to a version with only parallel (series) $LC$ resonators by using admittance (impedance) inverters, with an example in Fig.~\ref{fig1}(c) using admittance inverters and identical parallel $LC$ resonators \cite{Pozar2011}. Here we set the source and load impedance to be $Z_0=50~\mathrm{\Omega}$, typical for circuit QED, as these circuits are typically driven and measured by $50~\mathrm{\Omega}$ impedance-matched electronics. Furthermore, the admittance/impedance inverters can be treated as passive couplers between the resonant $LC$ elements \cite{Matthaei1980,Naaman2022}, from which we get the coupled-mode picture of a bandpass filter, as shown in Fig.~\ref{fig1}(d), where each red circle represents a resonator with frequency $\omega_0$, solid lines represent the coupling $c_{j,k}$ between the $j$th and $k$th resonators, and undulating arrows represent the dissipation $\kappa_j$ of the $j$th element; note we assume zero intrinsic loss for all the resonators. The coupling strength and dissipation rates are given in terms of the coefficients $g_j$ and bandwidth $\Delta \omega$ by \cite{Matthaei1980,Naaman2022}
\begin{align}
  c_{j,j+1} &= \frac{\Delta\omega}{2\sqrt{g_{j}g_{j+1}}},\\
  \kappa_1 &= \frac{\Delta\omega}{g_0g_1},\\
  \kappa_N &= \frac{\Delta\omega}{g_Ng_{N+1}}.
\end{align}

The transmission of two example $N=6$ order bandpass filters with center frequency $\omega_0/2\pi=6~\mathrm{GHz}$ and bandwidth $\Delta \omega/2\pi=600~\mathrm{MHz}$ are shown in Fig.~\ref{fig1}(e), showing flat frequency response in the bandpass as desired. From Table~\ref{tab:table1}, for $0~\mathrm{dB}$ insertion loss, the $g$ coefficients are symmetric and $\kappa_1=\kappa_N$, while for $20~\mathrm{dB}$ insertion loss, $\kappa_1 \ll \kappa_N$. We note that when the insertion loss goes to infinity, $g_jg_{j+1}$ ($j\geq 1$) saturates, while $\kappa_1=\Delta\omega/g_0g_1$ goes to zero, approaching the singly-terminated filter limit. (see supplementary material). The single-pole bandpass filter introduced in Ref.~\onlinecite{Jeffrey2014} can be treated as an $N=1$ bandpass filter with large insertion loss. Similar techniques have been used to design broadband Josephson parametric amplifiers \cite{Roy_2015,Grebel2021,White2023,Kaufman2023} and circulators \cite{Beck2022,Kwende2023}. We will label the port coupled to the first (last) filter stage as the input (output) port, where the output port is always strongly coupled to the environment to maximize collection of photons emitted from the readout resonator.

To realize qubit measurement, a qubit's readout resonator needs to be coupled to the filter. To better understand the differences between stages in the filter, and thus where to connect the readout resonator, we calculate the local density of states (LDOS) $\rho_{j}(\omega)$ at the $j$th filter stage. When the readout resonator is coupled to the $j$th filter stage, its dissipation rate $\kappa_r$ is proportional to the $j$th stage LDOS at $\omega_r$: $\kappa_r \propto \rho_j(\omega_r)$. Assuming the qubit frequency $\omega_q$ is in the stopband, the ratio $\rho_{j}(\omega_r)/\rho_j(\omega_q)$ quantifies how well the qubit is protected. The connection between the LDOS and classical circuit impedance is explained in the supplementary material. The LDOS is given by the diagonal elements of the imaginary part of the retarded Green's function $G_{jk}^R(\omega)$\cite{Harrison1980,Bruus2004},
\begin{align}
  G_{jk}^R(\omega) &= \int \mathrm{d}te^{i\omega t}G_{jk}^R(t)= -i\int \mathrm{d}te^{i\omega t}\theta(t)\langle[ \hat{a}_j(t),\hat{a}_k^\dagger(0)]\rangle,\\
  \rho_{j}(\omega) &= -\frac{1}{\pi}\mathrm{Im}G_{jj}^R(\omega),
\end{align}
where $\hat{a}_j^\dagger(t)$ ($\hat{a}_j(t)$) is the creation (annihilation) operator of an excitation in the $j$th filter stage in the Heisenberg picture, $\theta(t)$ is the Heaviside step function, and $\langle\cdot\rangle$ is the ground state expectation value.
The retarded Green's function $G^R_{jk}(\omega)$ measures the response of mode $j$ to a probe signal of frequency $\omega$ applied to mode $k$, with the imaginary part yielding the linear response susceptibility; the diagonal elements ${\rm Im}G^R_{jj}(\omega)$ thus describe the susceptibility of mode $j$ to a single of frequency $\omega$ impinging on it, and thus how easily a resonator of frequency $\omega$ will decay when coupled to that mode. While $G^R_{jk}(\omega)$ is just a linear response susceptibility \cite{clerk_Introduction}, and thus can be derived from classical coupled mode theory, we choose the quantum formalism here because the filter will ultimately couple to a qubit. The quantum theory directly connects to Fermi's golden rule, enabling direct calculation of qubit decay rates \cite{clerk_Introduction}.

\begin{figure}
  \includegraphics[]{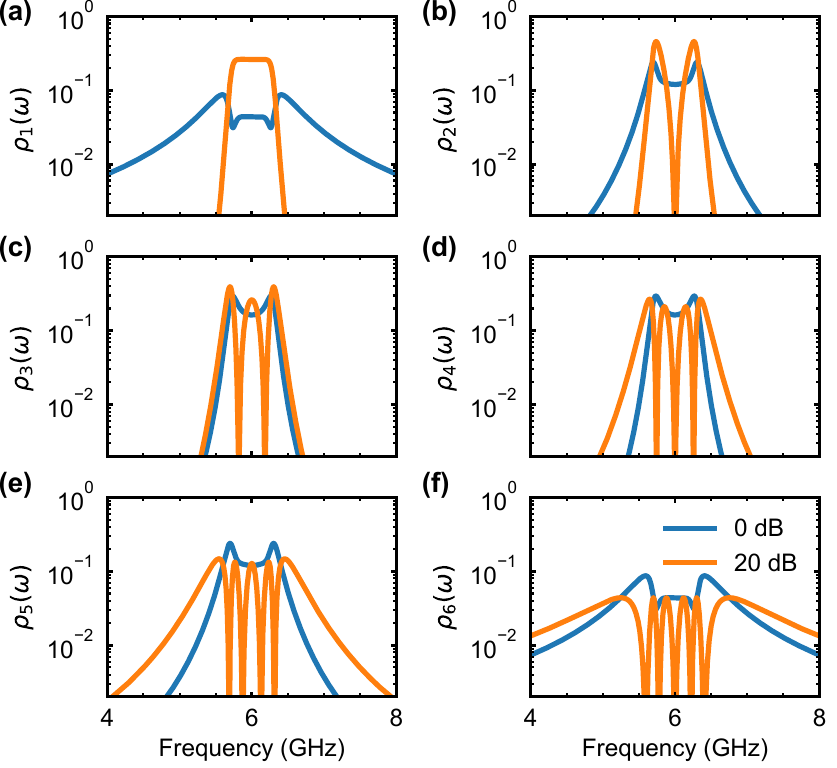}
  \caption{\label{fig2}(a-f) Local density of states (LDOS) $\rho_{j}(\omega)$ for the filter stages $j = 1$ to $6$ for $N=6$ order bandpass filters with $0~\mathrm{dB}$ (blue) and $20~\mathrm{dB}$ (orange) insertion loss. Bandpass filter center frequency $\omega_0/2\pi = 6~\mathrm{GHz}$ and bandwidth $\Delta \omega/2\pi = 600~\mathrm{MHz}$. We display the LDOS for $N=2$ to $5$ order filters in the supplementary material.}
\end{figure}

The calculated LDOS for $N=6$ bandpass filters with $0~\mathrm{dB}$ and $20~\mathrm{dB}$ insertion loss are shown in Fig.~\ref{fig2}. We can see for the $20~\mathrm{dB}$ insertion loss filter, $\rho_{1}(\omega)$ is flat in the passband and there are $j-1$ near-zero points in $\rho_{j}(\omega)$. When the readout resonator is coupled to the $j$th filter stage and its frequency is close to these near-zero points, it barely decays and cannot be used to do qubit readout (see supplementary material). We see similar features in larger insertion loss filters, which are caused by interference between different filter stages. For a symmetric filter, $\rho_j(\omega_q)$ is smallest when the readout resonator is coupled to either middle stage $j=3$ or $4$, meaning a qubit is better protected when its readout resonator is coupled to this point. In the following discussion, we will couple the readout resonators to the middle (first) stage of the symmetric $0~\mathrm{dB}$ (asymmetric $20~\mathrm{dB}$) insertion loss filter.

\begin{figure}
  \includegraphics[]{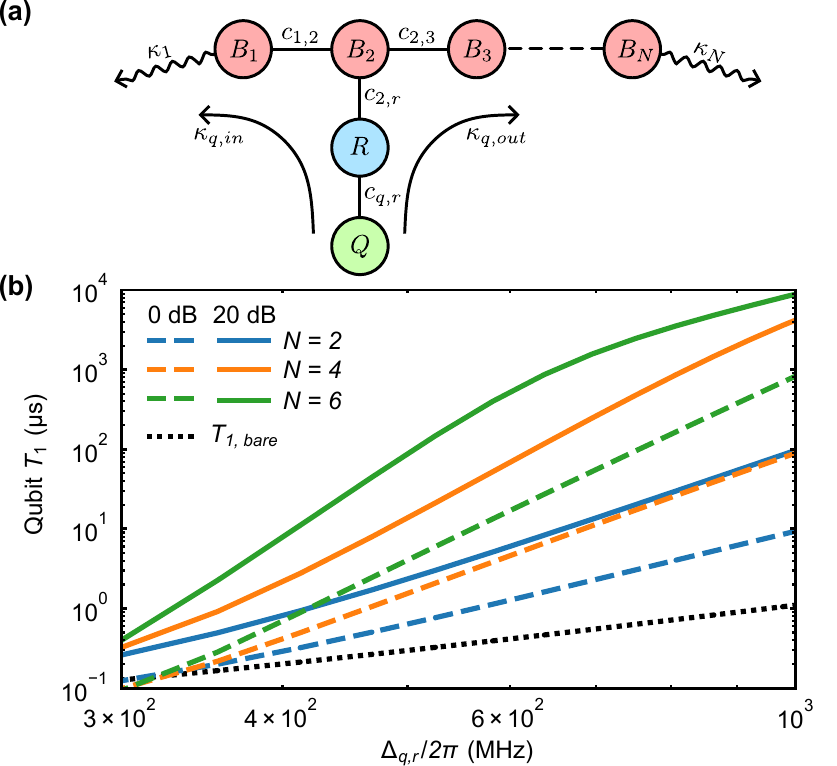}
  \caption{\label{fig3}(a) Coupled-mode picture of dispersive readout circuit, with qubit $Q$ coupled through readout resonator $R$ to filter element $B_j$. (b) Qubit lifetime $T_1$ versus qubit-resonator detuning $\Delta_{q,r}$ for bandpass filters with $0~\mathrm{dB}$ (dashed lines) and $20~\mathrm{dB}$ (solid lines) insertion loss. The filter center frequency $\omega_0/2\pi$ is $6~\mathrm{GHz}$, its bandwidth $\Delta \omega/2\pi$ is $600~\mathrm{MHz}$, the readout resonator frequency $\omega_r/2\pi$ is $6~\mathrm{GHz}$, the readout resonator dissipation rate $\kappa_r/2\pi$ is $15~\mathrm{MHz}$, and the qubit-resonator coupling $c_{q,r}/2\pi$ is $100~\mathrm{MHz}$.}
\end{figure}

The coupled-mode picture of the qubit readout circuit is shown in Fig.~\ref{fig3}(a), where $c_{q,r}$ is the qubit-readout resonator coupling strength and $c_{j,r}$ is the coupling strength between the readout resonator and the $j$th filter stage. The qubit can decay through the readout resonator and bandpass filter modes to the input (characterized by $\kappa_{q,in}$) and output ports (characterized by $\kappa_{q,out}$). When there is no Purcell filter, the qubit lifetime $T_{1,\mathrm{bare}}$ will be limited as in Ref.~\onlinecite{Purcell1995},
\begin{equation}
  T_{1,\mathrm{bare}}=\frac{\Delta_{q,r}^2}{\kappa_r c_{q,r}^2},
\end{equation}
where $\kappa_r$ is the readout resonator dissipation rate and $\Delta_{q,r}=\omega_q-\omega_r$ is the frequency detuning between the qubit and readout resonator. For $\kappa_r/2\pi=15~\mathrm{MHz}$ ($\kappa_r^{-1}\simeq 10~\mathrm{ns}$), $\Delta_{q,r}/2\pi=-1~\mathrm{GHz}$ ($\omega_q/2\pi=5~\mathrm{GHz}$) and $c_{q,r}/2\pi=100~\mathrm{MHz}$, the qubit relaxation time $T_1$ is limited to $1~\mathrm{\mu s}$. To quantify the protection from the bandpass filter, we display the qubit lifetime $T_1$ versus qubit-resonator detuning $\Delta_{q,r}$ with the filter as shown in Fig.~\ref{fig3}(b), where we treat the transmon qubit as a resonator and use the classical coupled-resonator model to extract the qubit $T_1$. Adding more stages and using larger qubit-resonator detuning gives longer qubit lifetime. Comparing a $0~\mathrm{dB}$ with a $20~\mathrm{dB}$ insertion loss filter, the higher insertion loss filter affords better protection. Note that for a large insertion loss filter, $\kappa_1\ll\kappa_N$, so the qubit mainly decays through the output port ($\kappa_{q,in}\ll\kappa_{q,out}$). We find that for symmetric filters ($0~\mathrm{dB}$ insertion loss) with order $N=2k$ or $N=2k-1$, the qubit lifetime $T_1$ scales as $T_1\propto \Delta_{q,r}^{2k+2}$. For asymmetric filters (large insertion loss) with order $N$, the qubit lifetime scales as $T_1\propto \Delta_{q,r}^{2N+2}$, consistent with the single-pole bandpass filter \cite{Jeffrey2014} $T_1\propto \Delta_{q,r}^{4}$. When increasing the number of stages, the qubit $T_1$ is lower than power-law scaling for the $20~\mathrm{dB}$ insertion loss filter when the qubit-resonator detuning is large, due to the finite $\kappa_1$ as here the readout resonator is coupled to the first stage. If we further increase the insertion loss, or simply reduce $\kappa_1$, the qubit $T_1$ will be closer to power-law scaling (see supplementary material). Here we demonstrate that by adding additional stages to the bandpass filter, we can achieve better qubit protection.

\begin{figure}
  \includegraphics[]{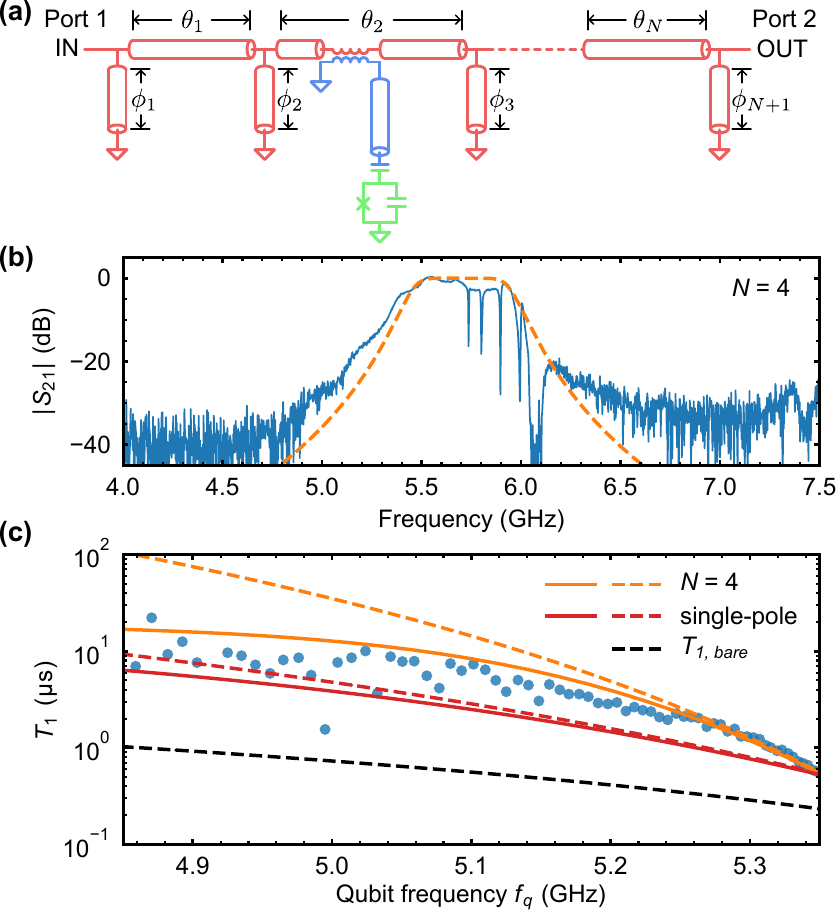}
  \caption{\label{fig4}Experimental realization. (a) Circuit diagram for an $N$th order bandpass filter implemented using sections of transmission line in place of lumped-element $LC$ resonators. (b) We experimentally implement a 4th order, zero insertion loss version of this circuit, coupling four qubits to the filter via their readout resonators, two connected to the second and two to the third filter stages. We measure the transmission $|S_{21}|$ of the qubit readout circuit, from input port (port 1) to output port (port 2); background attenuation is subtracted. There are four dips between $5.7$ to $6.0~\mathrm{GHz}$, corresponding to each of the four qubit readout resonators. The broader dip at $6.1~\mathrm{GHz}$ is due to the TWPA (see discussion). The orange dashed line is the coupled-mode simulation data for a symmetric 4th order bandpass filter with $5.7~\mathrm{GHz}$ center frequency and $500~\mathrm{MHz}$ bandwidth. (c) Qubit $T_1$ versus qubit frequency. The qubit is coupled to the $5.800~\mathrm{GHz}$ readout resonator. Dashed lines are simulated qubit $T_1$ limit with a 4th order symmetric filter (orange), a single-pole bandpass filter (red) and without any filter (black). Solid lines are qubit $T_1$ limit including the intrinsic qubit loss ($T_1=20~\mathrm{\mu s}$) with a 4th order symmetric filter (orange) and a single-pole bandpass filter (red). }
\end{figure}

In Fig.~\ref{fig4}a we display a simpler bandpass filter design, using transmission line resonators in place of lumped $L$ and $C$ elements, which can be more easily implemented in a thin-film planar geometry. We use shorted-to-ground transmission line elements to act as the impedance inverters \cite{Pozar2011}, where we adjust the length $\ell_n$ of each line in Fig.~\ref{fig4}(a) according to
\begin{align}
  \phi_n &= \arctan\left(\frac{Z_0 J_n}{1-(Z_0 J_n)^2}\right),\\
  \theta_n &= \pi-\arctan(Z_0 J_n)-\arctan(Z_0 J_{n+1}),
\end{align}
where $\phi_n$ and $\theta_n$ represent $\beta \ell_n$ with $\beta$ the phase constant at frequency $\omega_0$, and $J_n$ satisfies
\begin{align}
  Z_0 J_n &= \sqrt{\frac{\pi \Delta\omega}{2\omega_0g_{n-1}g_n}}\quad \text{for}~n=1,N+1,\\
  Z_0J_n &= \frac{\pi\Delta \omega}{2\omega_0\sqrt{g_{n-1}g_n}}\quad \text{for}~n=2,\cdots,N.
\end{align}
The figure shows a single qubit readout, where a quarter-wavelength transmission line resonator (blue) acts as the readout resonator, inductively coupled to the second bandpass filter stage and capacitively coupled to an Xmon qubit \cite{Koch2007,Barends2013} (green). More qubits can be read out by attaching them to the same or other filter stages; to couple more readout resonators to one filter stage, a $n\lambda/2$ transmission line resonator can be used.

The single-stage ($N=1$) asymmetric version of this circuit has been demonstrated in Refs.~\onlinecite{Bienfait2019,Zhong2021}. Compared to using capacitors as admittance inverters \cite{Pozar2011}, we only need to control the length of each line, which supports straightforward design and fabrication. Note that using transmission line elements means there will be other passbands at integer multiples of $\omega_0$ (see supplementary material). Other ways to implement similar bandpass filters include smaller footprint mechanical resonators \cite{Cleland2019,Chu2020} and lumped element $LC$ resonators \cite{Ferreira2022,Zhang2023,Scigliuzzo2022}. Some alternative circuit realizations are also shown in the supplementary material.

We experimentally implement a version of the circuit in Fig.~\ref{fig4}, using a symmetric (zero insertion loss), 4th order filter. We couple four qubits via their readout resonators to the bandpass filter. Two readout resonators are coupled to the second stage and the other two to the third stage of the filter. The circuit diagram of this sample is shown in the supplementary material.

The circuit is fabricated on two separate sapphire substrates. The qubits and readout resonators are fabricated on one die, and the bandpass filter and control wiring on a separate die. An aluminum base layer is first deposited by electron beam evaporation, and the circuit pattern defined by reactive plasma etching through a photoresist stencil. The qubit Josephson junctions are lift-off deposited using the Dolan bridge method \cite{Dolan1977}. The two dies are then aligned and attached to one another using a flip-chip bonding technique \cite{Satzinger2019,Conner2021}. 

The assembly is wirebonded into a chip mount and cooled to 10~mK on the mixing chamber stage of a dilution refrigerator. The filter output signal is amplified by a traveling-wave parametric amplifier (TWPA) \cite{Macklin2015} at the $10~\mathrm{mK}$ stage and a cryogenic HEMT amplifier at the $4~\mathrm{K}$ stage. The transmission $|S_{21}|$ of the readout circuit, measured from the input port (port 1) to the output port (port 2), is shown in Fig.~\ref{fig4}(b). The bandpass filter is designed to have a center frequency $\omega_0/2\pi=5.70~\mathrm{GHz}$, bandwidth $\Delta \omega/2\pi=500~\mathrm{MHz}$, and maximally flat response. The frequencies of the four readout resonators are $5.736$, $5.800$, $5.896$, and $5.993~\mathrm{GHz}$, as seen in the transmission data. These resonators have loaded quality factors $Q_l$ around $500$, corresponding to a dissipation rate $\kappa_r \approx 2\pi\times12~\mathrm{MHz}$. There are some ripples in the passband, which are possibly due to the flip-chip integration \cite{Li2023}. 

We measured the qubit relaxation time $T_1$ versus qubit frequency $f_q$, where the qubit is tuned by applying a rectangular current pulse to its flux-bias coil during the measurement. The measured results are shown in Fig.~\ref{fig4}(c). The experimental data (blue dots) at high frequencies agree well with simulation (orange dashed line) and are above the qubit $T_1$ limit with a single-pole bandpass filter \cite{Jeffrey2014} (red dashed line) and without the filter (black dashed line). The qubit lifetime at lower frequencies is shorter than the simulation results, which is limited by other loss mechanisms. If we consider the qubit intrinsic loss (i.e. $T_1$ is limited to $20~\mu\mathrm{s}$), the simulation data (orange solid line) agrees better with the experimental data. Similar results were found for the other three qubits. We also demonstrate multiplexed qubit readout, with the results shown in the supplementary material. We also fabricated and measured a second-order asymmetric filter, whose transmission data is shown in the supplementary material.

We have characterized our device using transmission from the input to the output port. Another common method is to perform a reflection measurement from the input port. In our scheme, this can be realized by using a large insertion loss bandpass filter and measuring the reflection from the output port (setting $\kappa_1$ to zero as a singly-terminated filter). We note that individual Purcell filters for each readout resonator can be added to suppress off-resonant driving \cite{Heinsoo2018}.

In summary, we present a systematic way to design and analyze broadband bandpass Purcell filters for circuit QED. We numerically show that large insertion loss filters can provide better qubit protection compared to conventional symmetrical filters. We experimentally implement these filters using a simple transmission line implementation and measure the performance for a 4th order symmetric filter implementation. This design can be easily integrated into existing superconducting quantum processor designs.

\section*{Supplementary Material}
In the Supplementary Material we provide additional details for constructing bandpass filters with different insertion loss values as well as alternative circuit implementations. We also provide more detailed explanations of the coupled-mode theory and local density of states. We also provide more details on the qubit experiments.

\section*{acknowledgments}
The authors want to thank Ofer Naaman for helpful and insightful comments. The authors also want to thank W. D. Oliver and G. Calusine at MIT Lincoln Lab for providing the traveling-wave parametric amplifier (TWPA) used in this work. This work was supported by the NSF QLCI for HQAN (NSF Award 2016136), by the Air Force Office of Scientific Research, and in part based on work supported by the U.S. Department of Energy Office of Science National Quantum Information Science Research Centers, and by UChicago's MRSEC (NSF award DMR-2011854). We made use of the Pritzker Nanofabrication Facility, which receives support from SHyNE, a node of the National Science Foundation's National Nanotechnology Coordinated Infrastructure (NSF Grant No. NNCI ECCS-2025633). 

\section*{References}
\bibliography{bpf_ref}
\end{document}



\title[]{Supplementary Information for ``Broadband Bandpass Purcell Filter for Circuit Quantum Electrodynamics''}
\author{Haoxiong Yan}
\affiliation{Pritzker School of Molecular Engineering, University of Chicago, Chicago IL 60637, USA}

\author{Xuntao Wu}
\affiliation{Pritzker School of Molecular Engineering, University of Chicago, Chicago IL 60637, USA}

\author{Andrew Lingenfelter}
\affiliation{Pritzker School of Molecular Engineering, University of Chicago, Chicago IL 60637, USA}
\affiliation{Department of Physics, University of Chicago, Chicago IL 60637, USA}

\author{Yash J. Joshi}
\affiliation{Pritzker School of Molecular Engineering, University of Chicago, Chicago IL 60637, USA}

\author{Gustav Andersson}
\affiliation{Pritzker School of Molecular Engineering, University of Chicago, Chicago IL 60637, USA}

\author{Christopher R. Conner}
\affiliation{Pritzker School of Molecular Engineering, University of Chicago, Chicago IL 60637, USA}

\author{Ming-Han Chou}
\affiliation{Pritzker School of Molecular Engineering, University of Chicago, Chicago IL 60637, USA}
\affiliation{Department of Physics, University of Chicago, Chicago IL 60637, USA}

\author{Joel Grebel}
\affiliation{Pritzker School of Molecular Engineering, University of Chicago, Chicago IL 60637, USA}

\author{Jacob M. Miller}
\affiliation{Pritzker School of Molecular Engineering, University of Chicago, Chicago IL 60637, USA}
\affiliation{Department of Physics, University of Chicago, Chicago IL 60637, USA}

\author{Rhys G. Povey}
\affiliation{Pritzker School of Molecular Engineering, University of Chicago, Chicago IL 60637, USA}
\affiliation{Department of Physics, University of Chicago, Chicago IL 60637, USA}

\author{Hong Qiao}
\affiliation{Pritzker School of Molecular Engineering, University of Chicago, Chicago IL 60637, USA}

\author{Aashish A. Clerk}
\affiliation{Pritzker School of Molecular Engineering, University of Chicago, Chicago IL 60637, USA}

\author{Andrew N. Cleland}
\affiliation{Pritzker School of Molecular Engineering, University of Chicago, Chicago IL 60637, USA}
\affiliation{Argonne National Laboratory, Lemont IL 60439, USA}
\date{\today}

\maketitle
\setcounter{figure}{0}
\setcounter{equation}{0}
\setcounter{table}{0}
\renewcommand{\thefigure}{S\arabic{figure}}
\renewcommand{\thetable}{S\arabic{table}}
\renewcommand{\theequation}{S\arabic{equation}}

\section{$g$ coefficients}
In Table \ref{tab:tableS1} we list the $g$ coefficients for maximally flat, low-pass filter designs with $10$ and $30~\mathrm{dB}$ insertion loss. We can see that $g_jg_{j+1}$ for $20~\mathrm{dB}$ (main text) and $30~\mathrm{dB}$ insertion loss filters are very close when $j\geq 1$, and differ by very close to a factor of 10 for $j = 1$.
\begin{table*}[b]
  \caption{\label{tab:tableS1}$g$ coefficients for maximally flat low-pass filter prototypes, with $g_0=1$.}
  \begin{ruledtabular}
  \begin{tabular}{ccccccccc}
  Insertion loss&Order&$g_0g_1$&$g_1g_2$&$g_2g_3$&$g_3g_4$&$g_4g_5$&$g_5g_6$&$g_6g_7$\\
  \hline
  10 dB& 1 &38.97&1.026&&&&\\
   & 2 &54.40&1.026&0.716&&&&\\
   & 3 &57.45&2.035&0.678&0.504&&&\\
   & 4 &58.50&2.445&1.730&0.420&0.385&&\\
   & 5 &58.97&2.645&2.366&1.249&0.279&0.311&\\
   & 6 &59.22&2.755&2.755&1.882&0.919&0.198&0.260\\
   \hline
   30 dB& 1 &3999&1.000&&&&&\\
   & 2 &5655&1.000&0.707&&&&\\
   & 3 &5997&2.000&0.667&0.500&&&\\
   & 4 &6120&2.415&1.707&0.414&0.383&&\\
   & 5 &6178&2.618&2.342&1.236&0.276&0.309&\\
   & 6 &6209&2.732&2.732&1.866&0.911&0.196&0.259\\
  \end{tabular}
  \end{ruledtabular}
  \end{table*}

\section{Coupled-mode picture}
\begin{figure*}
  \includegraphics[width=\textwidth]{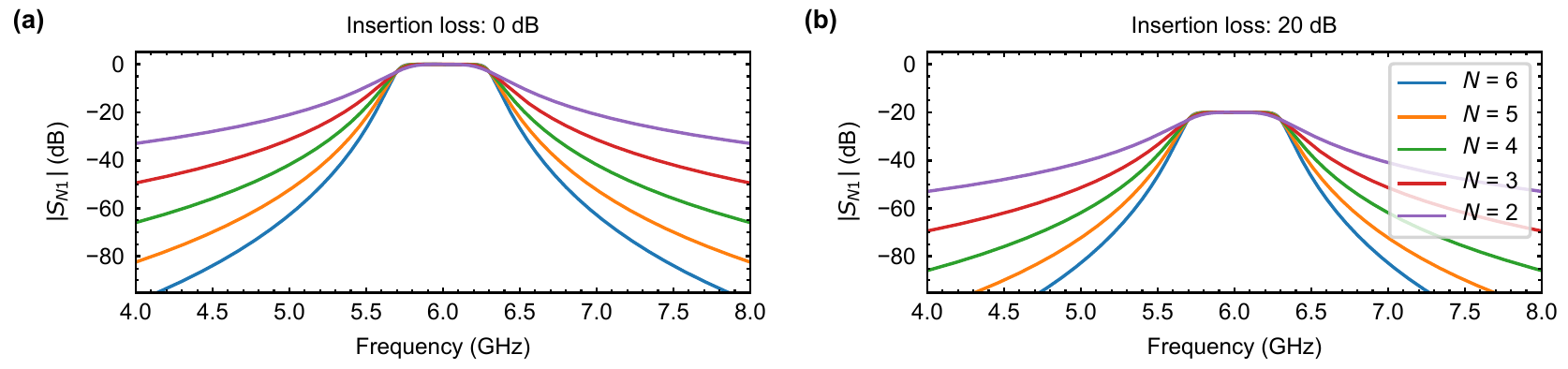}
  \caption{Transmission coefficient $|S_{N1}|$ for different filter stage numbers with insertion loss (a) $0~\mathrm{dB}$ and (b) $20~\mathrm{dB}$. All the filters have the same center frequency $\omega_0/2\pi=6~\mathrm{GHz}$ and bandwidth $\Delta \omega/2\pi =600~\mathrm{MHz}$.}\label{figS1}
\end{figure*}

The multi-stage bandpass filter introduced in the main text can be treated as a chain of coupled harmonic oscillators (resonators). The Hamiltonian $\hat{H}$ can be written as \cite{Naaman2022}
\begin{equation}\label{equ:coupled mode}
  \hat{H}/\hbar=\sum_{j=1}^N\omega_0 \hat{a}_j^\dagger \hat{a}_j+\sum_{j=1}^{N-1}c_{j,j+1}(\hat{a}_j \hat{a}_{j+1}^\dagger+\hat{a}_j^\dagger \hat{a}_{j+1})
\end{equation}
where $\hat{a}_j^\dagger$ ($\hat{a}_j$) is the creation (annihilation) operator of an excitation in the $j$th resonator, $\omega_0$ is the filter center frequency, $c_{j,j+1}$ is the coupling strength between the $j$th and $(j+1)$th resonators, and the rotating wave approximation (RWA) has been applied. The Heisenberg equation of motion for $\vec{A}(t)=[\hat{a}_1(t),\hat{a}_2(t),\cdots, \hat{a}_{N}(t)]^{T}$ is
\begin{align}
  \frac{\mathrm{d}}{\mathrm{d}t}\vec{A}(t)&=\mathbf{M}\vec{A}(t)-\mathbf{K}\vec{A}_{in}(t),\label{equ:EOM}\\
  \mathbf{M}_{jk}&=-(i\omega_0+\frac{\kappa_j}{2})\delta_{jk}-ic_{j,j+1}(\delta_{j+1,k}+\delta_{j-1,k}),\\
  \mathbf{K}_{jk}&=\delta_{jk}\sqrt{\kappa_j},
\end{align}
where $\vec{A}_{in}=[\hat{a}_{1}^{in}(t),\hat{a}_{2}^{in}(t),\cdots, \hat{a}_{N}^{in}(t)]^{T}$ is the vector of input fields, $\kappa_j$ is the external dissipation rate of $j$th resonator, and we assume the internal dissipation rates are zero for all the resonators. We have the input-output relation \cite{Gardiner1985}
\begin{equation}
  \vec{A}_{out}(t)=\mathbf{K}\vec{A}(t)+\vec{A}_{in}(t),\label{equ:input-output}
\end{equation}
where $\vec{A}_{out}=[\hat{a}_{1}^{out}(t),\hat{a}_{2}^{out}(t),\cdots, \hat{a}_{N}^{out}(t)]^{T}$ is the vector of output fields. Assuming we drive the resonators with signal of frequency $\omega$ and calculating the Fourier transforms of Eq. \ref{equ:EOM} and \ref{equ:input-output}, we have
\begin{align}
  -i\omega \vec{A}[\omega]&=\mathbf{M}\vec{A}[\omega]-\mathbf{K}\vec{A}_{in}[\omega],\\
  \vec{A}_{out}[\omega]&=\mathbf{K}\vec{A}[\omega]+\vec{A}_{in}[\omega],
\end{align}
from which we can get the transmission coefficient $S_{N1}$,
\begin{equation}
  S_{N1}[\omega]=\frac{\hat{a}_{N,out}[\omega]}{\hat{a}_{1,in}[\omega]}=\sqrt{\kappa_1\kappa_N}\big[(i\omega\mathbf{I}+\mathbf{M})^{-1}\big]_{N1}.
\end{equation}
In Fig.~\ref{figS1} we display the transmission coefficient $S_{N1}$ for different filter stage numbers and insertion loss values.

\section{Local density of states}
\begin{figure*}
  \includegraphics[width=\textwidth]{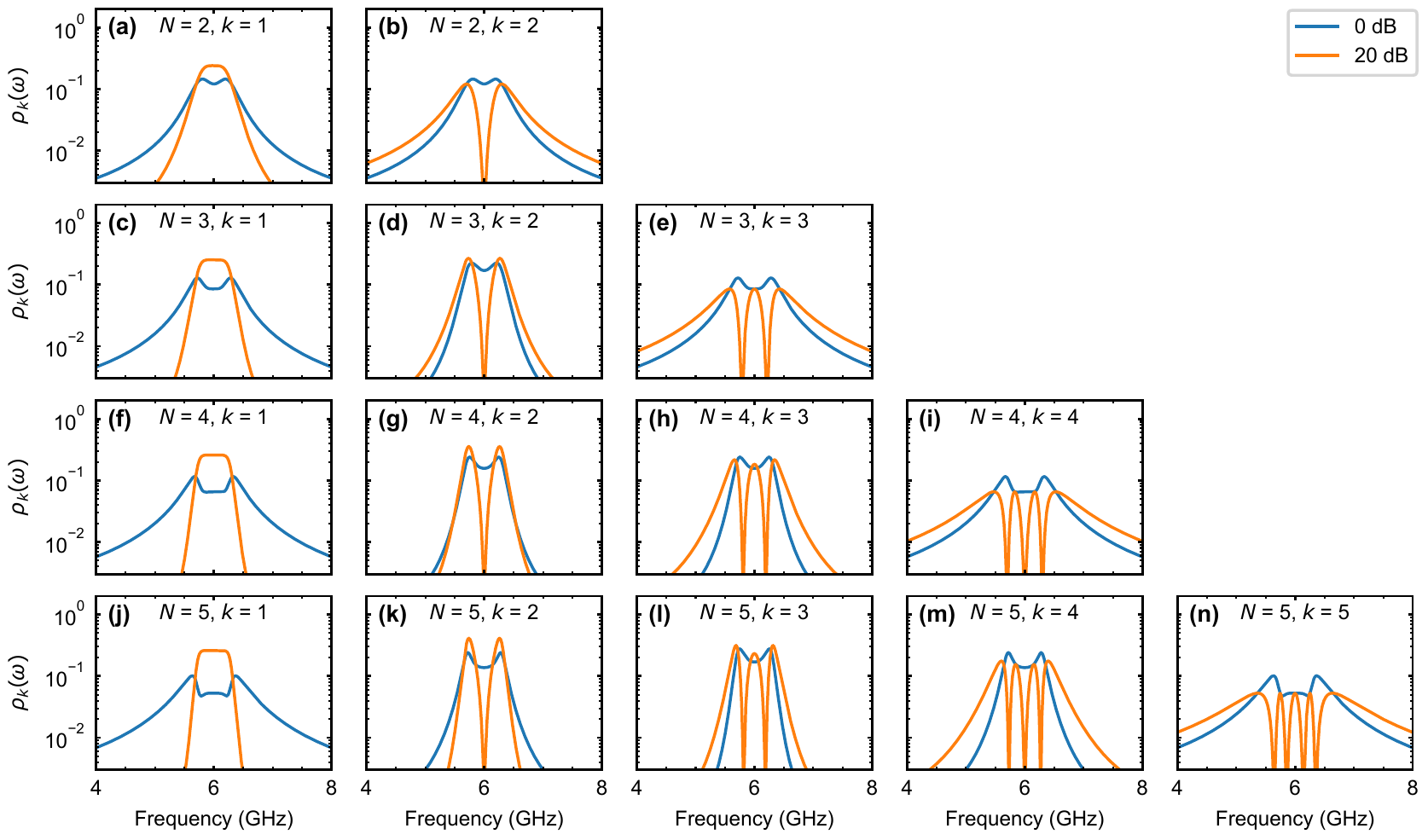}
  \caption{Local density of states (LDOS) $\rho_{k}(\omega)$ for different filter stage numbers $k$ for $N$th-order bandpass filters with $0~\mathrm{dB}$ (blue) and $20~\mathrm{dB}$ (orange) insertion loss. Bandpass filter center frequency $\omega_0/2\pi = 6~\mathrm{GHz}$ and bandwidth $\Delta \omega/2\pi = 600~\mathrm{MHz}$.}\label{figS2}
\end{figure*}

We calculate the local density of states (LDOS) using linear response theory. For readers who are not familiar with quantum linear response theory \cite{clerk_Introduction}, here we give a short introduction. Consider a system with Hamiltonian $H$ applied with some external field $H_{ext}=f(t)B$ where $B$ is an operator in the system, the total Hamiltonian is $H_{tot}=H+H_{ext}$. For the operator $A$, we have
\begin{align}\label{equ:linear response}
    \delta\langle A(t)\rangle = -i\int_{-\infty}^t\mathrm{d}t'\langle [A(t), B(t')]\rangle f(t'),
\end{align}
where $\delta\langle A(t)\rangle$ is the difference of $\langle A(t)\rangle$ with and without the external field and we keep terms to first order in $f(t)$. Eq.~\ref{equ:linear response} can be written as
\begin{align}
    \delta\langle A(t)\rangle =\int_{-\infty}^{\infty}\mathrm{d}t'G_{AB}^R(t-t') f(t'),
\end{align}
where the Green's function $G_{AB}^R(t-t')$ is given by 
\begin{align}
    G_{AB}^R(t-t')=-i\theta(t-t')\langle [A(t),B(t')]\rangle.
\end{align}
Here $\theta(t)$ is the Heaviside step function. The retarded Green's function thus gives the system's response to an external field.

For the coupled-mode model in Eq.~\ref{equ:coupled mode}, setting $A=\hat{a}_j$ and $B=\hat{a}^\dagger_k$, we have
\begin{align}
    G_{jk}^R(t)=-i\theta (t)\langle [\hat{a}_j (t),\hat{a}_k^\dagger(0)]\rangle.
\end{align}
The local density of states $\rho_j(\omega)$ (LDOS) is defined by the imaginary part of the Fourier transform of $G_{jk}^R(t)$,
\begin{align}
  G_{jk}^R(\omega) &= \int_{-\infty}^{+\infty}\mathrm{d}t e^{i\omega t}G_{jk}^R(t) = -i\int_{0}^{+\infty}\mathrm{d}t e^{i\omega t}\langle [\hat{a}_j (t),\hat{a}_k^\dagger(0)]\rangle,\\
  \rho_k(\omega) &= -\frac{1}{\pi}\mathrm{Im}G_{kk}^R(\omega),
\end{align}
which gives the decay rate of mode $k$ at frequency $\omega$. If a resonator with frequency $\omega_r$ is coupled to mode $k$, the resonator's decay rate $\kappa_r$ will thus be proportional to $\rho_k(\omega_r)$. Here $\{\hat{a}^\dagger_j(t)\}$ can be directly calculated from the equations of motion Eq. \ref{equ:EOM}, where the input fields $\{\hat{a}_j^{in}(t)\}$ are in their quantum ground states (at zero temperature). The calculated LDOS for $N=2$ to $5$ bandpass filters with $0~\mathrm{dB}$ and $20~\mathrm{dB}$ insertion loss are shown in Fig.~\ref{figS2}. For the $20~\mathrm{dB}$ insertion loss filters, $\rho_1(\omega)$ is flat in the passband and there are $k-1$ near-zero points in $\rho_k(\omega)$.

The connection between the LDOS and the classical circuit impedance can be understood from the following: $\rho_k(\omega_r)$ quantifies the dissipation rate $\kappa_r$ of a resonator with frequency $\omega_r$ when coupled to the $k$th stage of the filter: $\kappa_r\propto \rho_k(\omega_r)$. For a classical circuit, if we consider a parallel $LC$ resonator coupled to an external electrical circuit with impedance $Z_e(\omega)$ through a small capacitor $C_g$, its dissipation rate $\kappa_r$ is determined by the real part of the admittance $Y(\omega)$ seen by the resonator, $\kappa_r=\mathrm{Re}[Y(\omega_r)]/C$, where $C$ is the capacitance of the resonator. $Y(\omega)$ can be calculated as\cite{Blais2021}
\begin{align}
    Y(\omega) = \frac{i\omega C_g}{1+i\omega C_g Z_e(\omega)},\\
    \mathrm{Re[Y(\omega)]}=\omega^2 C_g^2\mathrm{Re}[Z_e(\omega)].
\end{align}
The LDOS is thus related to the circuit impedance as $\rho_k(\omega)\propto \mathrm{Re}[Z_e(\omega)]$, where $Z_e(\omega)$ is the circuit impedance from the resonator coupling point at the filter to ground. The ratio $\rho_k(\omega_r)/\rho_k(\omega_q)\propto \mathrm{Re}[Z_e(\omega_r)]/\mathrm{Re}[Z_e(\omega_q)]$ quantifies the qubit lifetime enhancement from the filter, which is consistent with Eq.~(3) derived in  Ref.~\onlinecite{cleland2019}.

If we consider a series $LC$ resonator inductively coupled to an external electrical circuit with impedance $Z_e(\omega)$ by a small mutual inductance $M$ (e.g. the circuit shown in Fig.~4(a) in the main text), its dissipation rate $\kappa_r$ is determined by the real part of the impedance $Z(\omega)$ seen by the resonator, $\kappa_r=\mathrm{Re}[Z(\omega_r)]/L$, where $L$ is the inductance of the resonator. $Z(\omega)$ can be calculated using
\begin{align}
    Z(\omega)=\frac{\omega^2 M^2}{Z_e(\omega)}=\omega^2 M^2 Y_e(\omega),\\
    \mathrm{Re}[Z(\omega)]=\omega^2 M^2\mathrm{Re}[Y_e(\omega)].
\end{align}
In this case, the LDOS is related to the circuit admittance as $\rho_k(\omega)\propto \mathrm{Re}[Y_e(\omega)]$, where $Y_e(\omega) $ is the circuit admittance from the resonator coupling point at the filter to ground. Here we choose to calculate the LDOS instead of the circuit impedance, because the circuit impedance depends on the circuit realization.

\begin{figure*}
  \includegraphics[width=\textwidth]{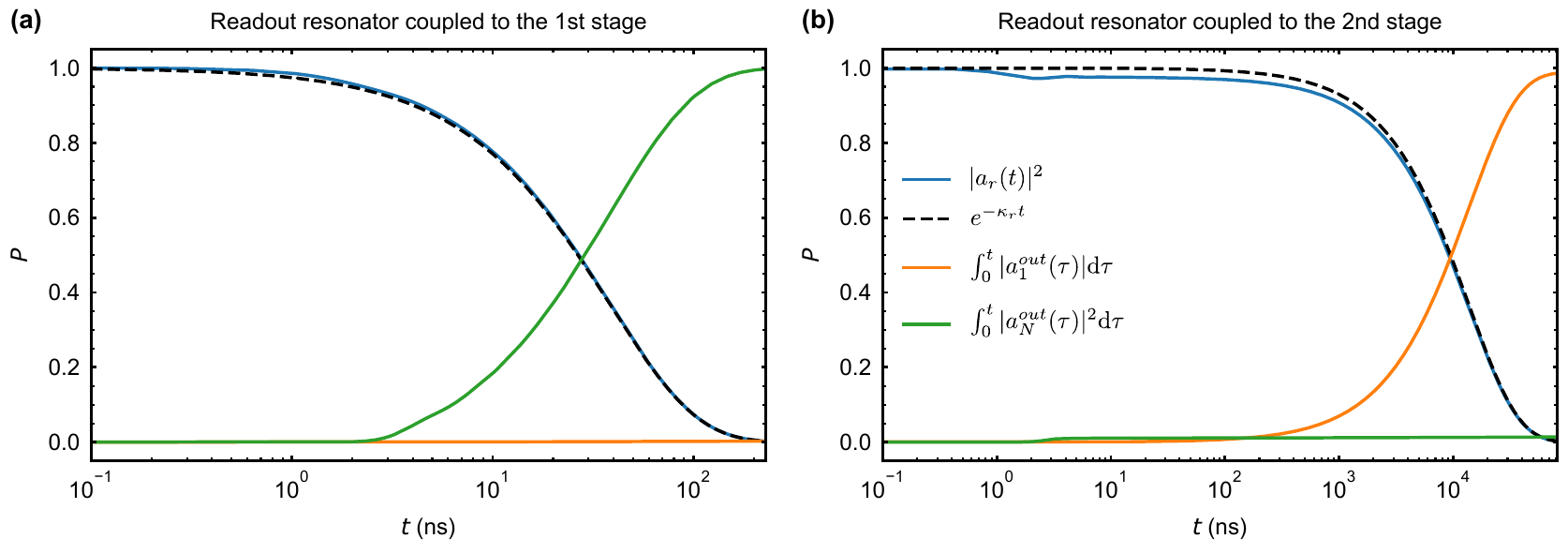}
  \caption{Evolution of an excitation $|a_r(t)|^2$ in a readout resonator coupled to (a) the 1st stage and (b) 2nd stage of a 6th order filter (blue lines), fit to an exponential decay (black dashed lines). We also show the photon numbers emitted from the input (orange) and output (green) ports. The coupling strength between the readout resonator and filter stage is $c_{r,k}/2\pi=20~\mathrm{MHz}$, the readout resonator frequency $\omega_r/2\pi=6~\mathrm{GHz}$, the filter center frequency $\omega_0/2\pi = 6~\mathrm{GHz}$, and the filter bandwidth $\Delta \omega/2\pi = 600~\mathrm{MHz}$.}\label{figS3}
\end{figure*}

We simulate the evolution of an excitation in a readout resonator when coupled to different filter stages. The results are shown in Fig.~\ref{figS3}, where a readout resonator is coupled to the first and second stage of a $6$th order, $20~\mathrm{dB}$ insertion loss filter. The coupling strength $c_{r,k}$ between the readout resonator and the $k$th filter stage is $2\pi\times 20~\mathrm{MHz}$, and the readout resonator frequency $\omega_r$ is equal to the filter center frequency $\omega_0$. We see that the readout resonator decays more slowly when it is coupled to the second stage, consistent with the LDOS calculation. We further calculate the photon leak rates from the input (orange) and output (green) ports. When the readout resonator is coupled to the first stage, most of the excitation energy leaks through the output port as expected, while when coupled to the second stage, most of the excitation leaks through the input port. Thus, in this filter design, the readout resonator cannot be coupled to the second stage, because its frequency is close to the near-zero point in the LDOS, and because its excitation energy leaks through the wrong filter port.

\section{Qubit $T_1$ scaling}
\begin{figure*}
  \includegraphics[width=\textwidth]{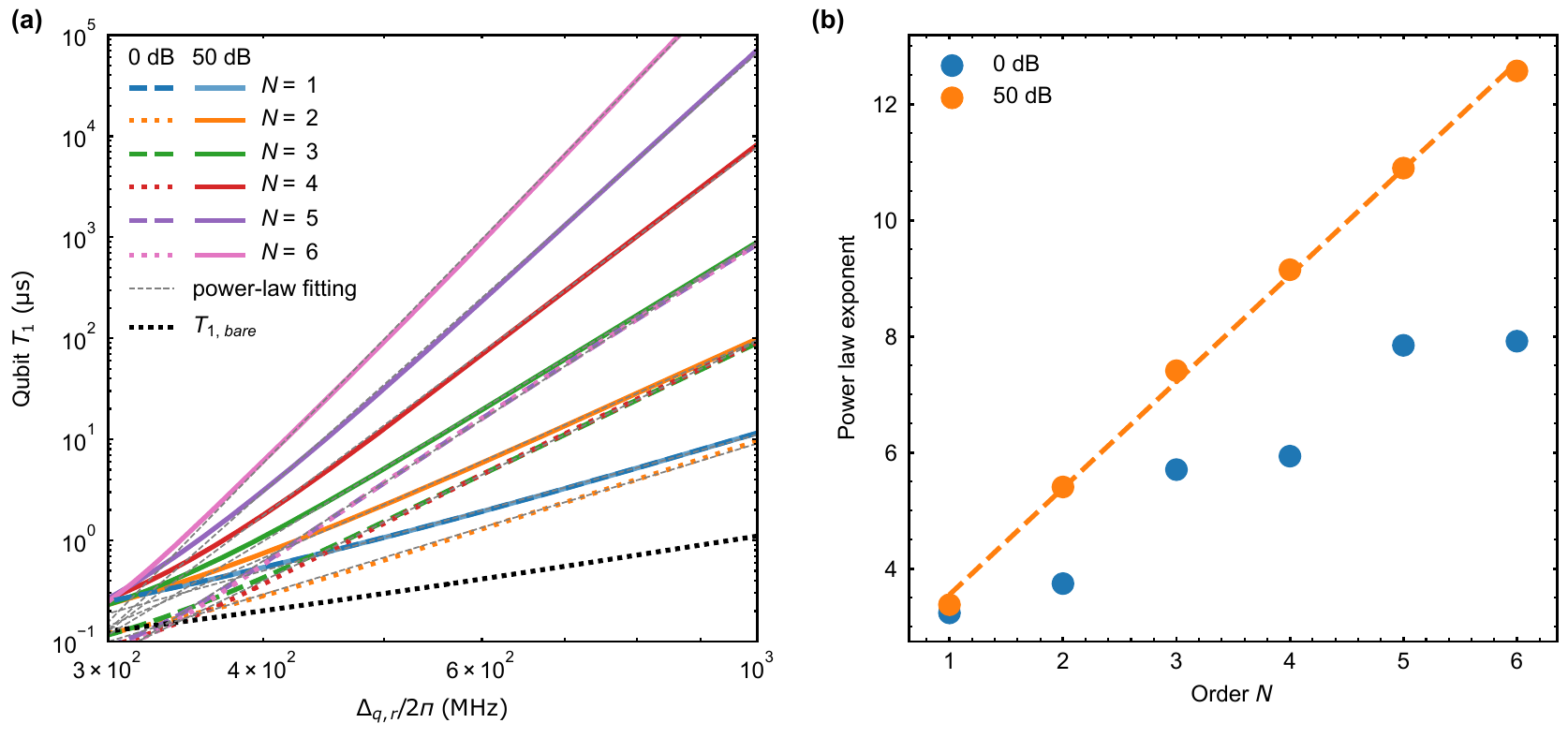}
  \caption{(a) Qubit lifetime $T_1$ versus qubit-resonator detuning $\Delta_{q,r}$ for the bandpass filters with $0~\mathrm{dB}$ (dashed and dotted lines) and $50~\mathrm{dB}$ (solid lines) insertion loss. Gray dashed lines are power-law fitting lines and the black dotted line is the qubit $T_1$ limited by Purcell effect $T_{1,\mathrm{bare}}$. (b) Fitted power law exponent versus filter order $N$ for $0~\mathrm{dB}$ (blue) and $50~\mathrm{dB}$ (orange) insertion loss. The filter center frequency $\omega_0/2\pi$ is $6~\mathrm{GHz}$, its bandwidth $\Delta \omega/2\pi$ is $600~\mathrm{MHz}$, the readout resonator frequency $\omega_r/2\pi$ is $6~\mathrm{GHz}$, the readout resonator dissipation rate $\kappa_r/2\pi$ is $15~\mathrm{MHz}$, and the qubit-resonator coupling $c_{q,r}/2\pi$ is $100~\mathrm{MHz}$}.\label{figS4}
\end{figure*}

A classical coupled-resonator model is used to extract qubit $T_1$ in Fig.~3(b) in the main text, where we model the transmon qubit as a resonator and calculate its decay rate. The RWA is not used in this calculation. In Fig.~\ref{figS4}(a), we plot the qubit $T_1$ versus qubit-resonator detuning $\Delta_{q,r}$ for filters with $0~\mathrm{dB}$ (dashed and dotted lines) and $50~\mathrm{dB}$ (solid lines) insertion loss. Compared with the $20~\mathrm{dB}$ insertion loss filter (main text), the qubit $T_1$ shows power-law scaling for larger insertion loss, which confirms that when $N$ and $|\Delta_{q,r}|$ is large, the qubit $T_1$ is limited by the non-zero $\kappa_1$. For symmetric filters, filters with order $N=2k$ (dashed lines) and $N=2k-1$ (dotted lines) have similar performance. We do the power-law fitting $T_1=a\cdot\Delta_{q,r}^b$ (gray dashed lines) on all the lines and the fitted exponents $b$ are shown in Fig.~\ref{figS4}(b). For asymmetric filters ($50~\mathrm{dB}$ insertion loss), We numerically find that the qubit $T_1$ scales approximately as $T_1\propto \Delta_{q,r}^{1.83N+1.72}$. We empirically find that these power law exponents are approximately two times the number of resonators the qubit decays through.

\section{Experimental realization with transmission line sections}
\begin{figure*}
  \includegraphics{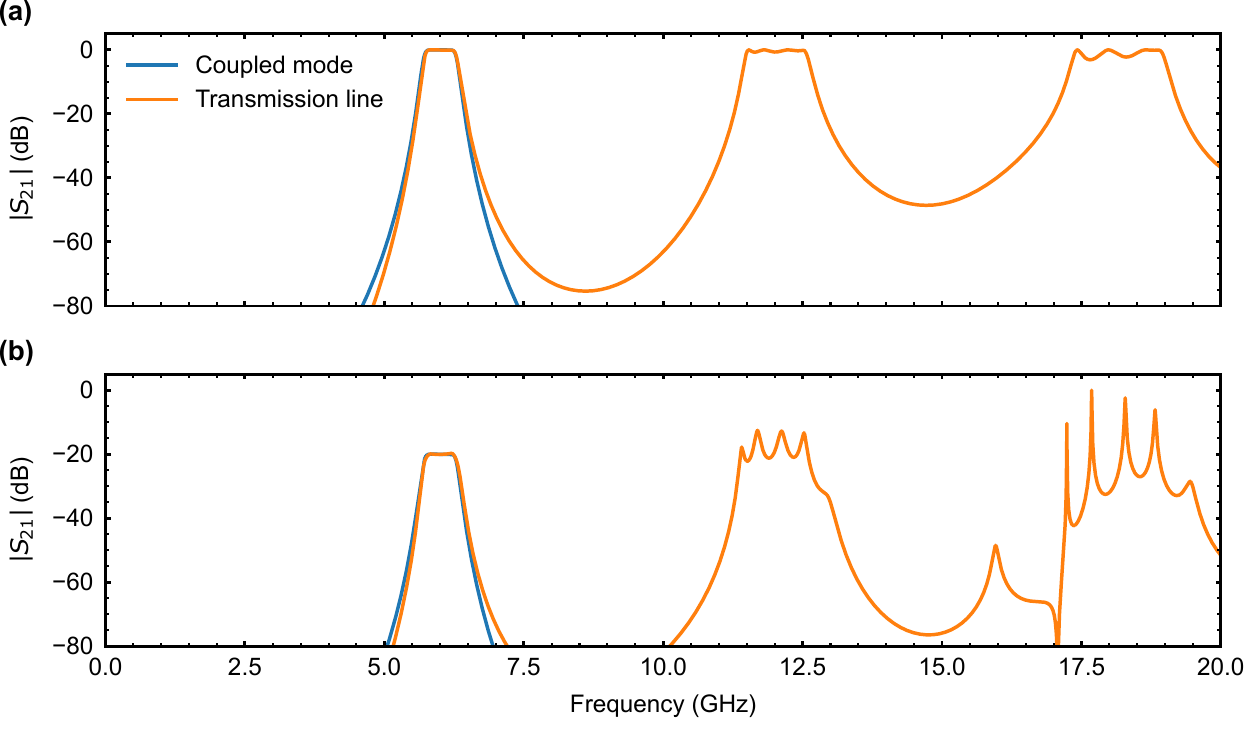}
  \caption{Transmission coefficient $|S_{21}|$ from the input port (port 1) to output port (port 2) for 6th-order bandpass filters using the coupled-mode picture (blue) and transmission-line-section realization (orange) with (a) $0~\mathrm{dB}$ and (b) $20~\mathrm{dB}$ insertion loss, center frequency $\omega_0/2\pi=6~\mathrm{GHz}$ and bandwidth $\Delta \omega/2\pi =600~\mathrm{MHz}$.}\label{figS5}
\end{figure*}

In Fig.~\ref{figS5}, we calculate the transmission coefficient $S_{21}$ for the bandpass filters realized with transmission line sections introduced in the main text. We can see other passbands at higher frequencies.

\begin{figure*}
  \includegraphics[width=0.75\textwidth]{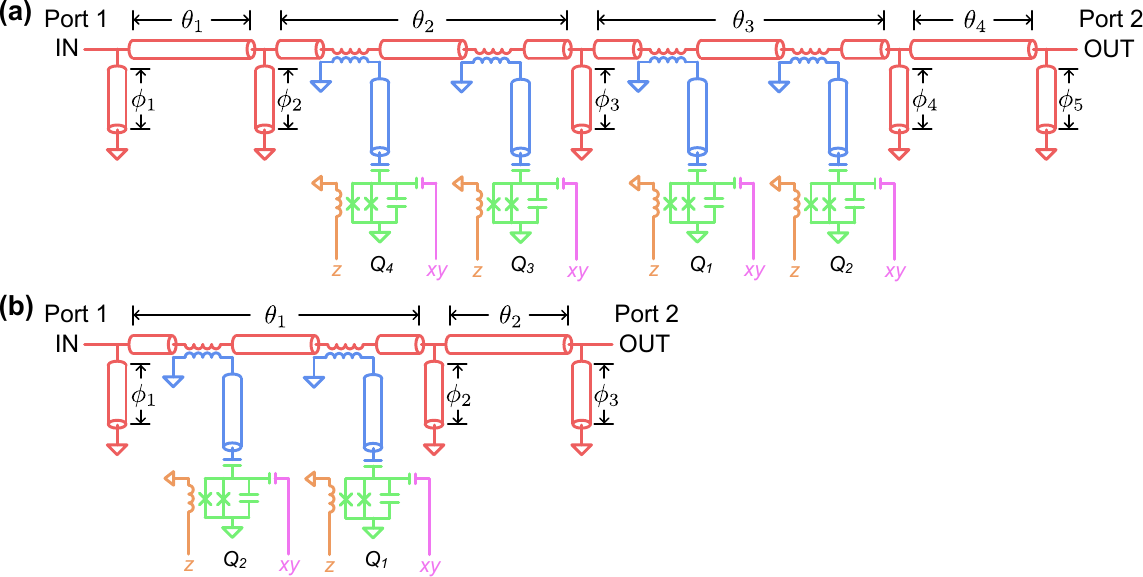}
  \caption{(a) Circuit diagram of the sample measured in Fig.~4(b) and (c) in the main text, with four qubits coupled to the second and third stages of a 4th order symmetric filter. (b) Circuit diagram of the sample measured in Fig.~\ref{figS8}, with two qubits coupled to the first stage of a second-order asymmetric filter. The qubits are labeled by the order of their readout resonators' frequencies, from low to high.}\label{figS6}
\end{figure*}

In Fig.~\ref{figS6}(a), we display the circuit diagram of the sample measured in Fig.~4(b) and (c) in the main text. There are four readout resonators coupled to an $N=4$ symmetric filter, two of which are coupled to the second stage, with the other two coupled to the third stage.

\begin{figure*}
  \includegraphics{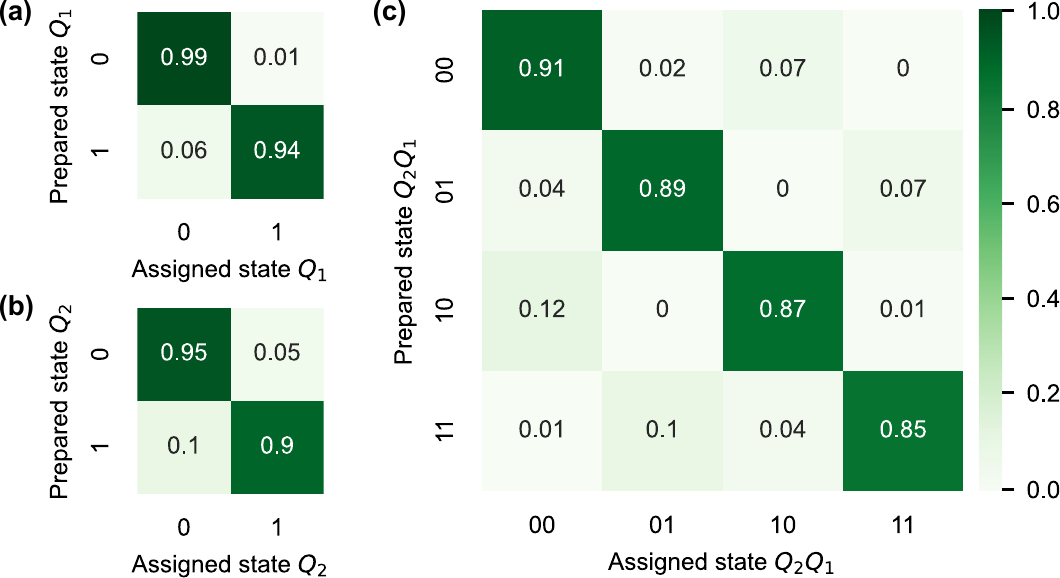}
  \caption{Assignment probability matrices $P(i|j)$ for (a) $Q_1$ single-qubit readout, (b) $Q_2$ single-qubit readout and (c) $Q_1$ and $Q_2$ multiplexed readout. $Q_1$ and $Q_2$ operate at $5.18~\mathrm{GHz}$, and are coupled to the $5.736~\mathrm{GHz}$ and $5.800~\mathrm{GHz}$ readout resonators, respectively. The coupling strength $c_{q,r}/2\pi$ between the qubit and readout resonator is around $110~\mathrm{MHz}$.}\label{figS7}
\end{figure*}

To show the multiplexed readout capacity, the readout assignment probability matrices of $Q_1$ and $Q_2$ are shown in Fig.~\ref{figS7}. The assignment probability matrix $P(i|j)$ is defined as the probability of assigning a qubit to state $j$ when preparing it in state $i$. For single qubit readout, the assignment probability matrices $P_{Q_1}(i|j)$ and $P_{Q_2}(i|j)$ are shown in Fig.~\ref{figS7}(a) and (b). The dominant readout errors come from qubit decay. For multiplexed readout, the assignment probability matrix $P_{Q_2Q_1}(i|j)$ is shown in Fig.~\ref{figS7}(c).

\begin{figure*}
  \includegraphics[width=\textwidth]{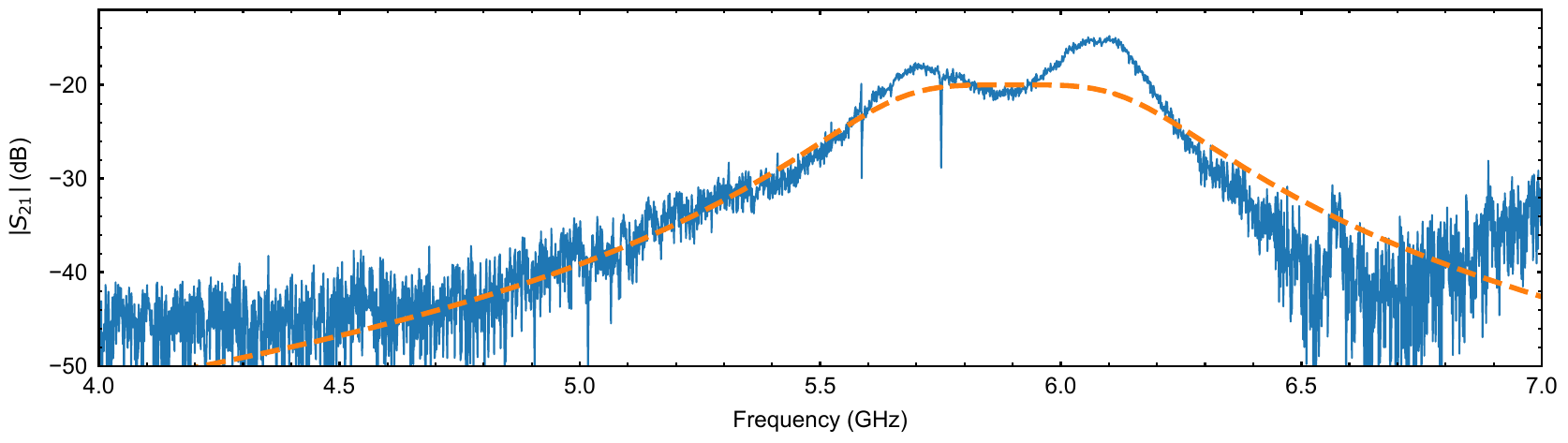}
  \caption{Transmission $|S_{21}|$ of the qubit readout circuit with a second-order asymmetric bandpass filter, from input port (port 1) to output port (port 2); background attenuation is subtracted. There are two dips at $5.586$ and $5.752~\mathrm{GHz}$, corresponding to each of the two qubit readout resonators. The orange dashed line is the simulation data for an asymmetric ($20~\mathrm{dB}$ insertion loss) second-order bandpass filter with $5.9~\mathrm{GHz}$ center frequency and $600~\mathrm{MHz}$ bandwidth.}\label{figS8}
\end{figure*}

We also experimentally implement a second-order, $20~\mathrm{dB}$ insertion loss version of the qubit readout circuit. The circuit is shown in Fig.~\ref{figS6}(b), where two readout resonators are coupled to the first stage of the asymmetric filter. The sample was fabricated following the same process as the sample shown in the main text, and cooled down to $100~\mathrm{mK}$ in an adiabatic demagnetization refrigerator (ADR) to measure its transmission coefficient. The results are shown in Fig.~\ref{figS8}. There are two readout resonators coupled to the first stage of the filter, with frequencies $5.586$ and $5.752~\mathrm{GHz}$ and coupling quality factors $Q_c$ around $8000$ and $2500$. The higher $Q_c$ for the $5.586~\mathrm{GHz}$ resonator is mainly because the resonator is outside the filter passband thus has weaker coupling, as shown in the LDOS calculation. There are large ripples in the passband, which is possibly caused by the resonator frequency shift due to the flip-chip integration. The larger transmission at frequencies above $7~\mathrm{GHz}$ is possibly due to a packaging mode. We are not able to measure the qubit lifetime, as this cryostat does not support this kind of measurement.

\section{Alternative experimental realizations}
\begin{figure*}[t]
    \includegraphics[width=\textwidth]{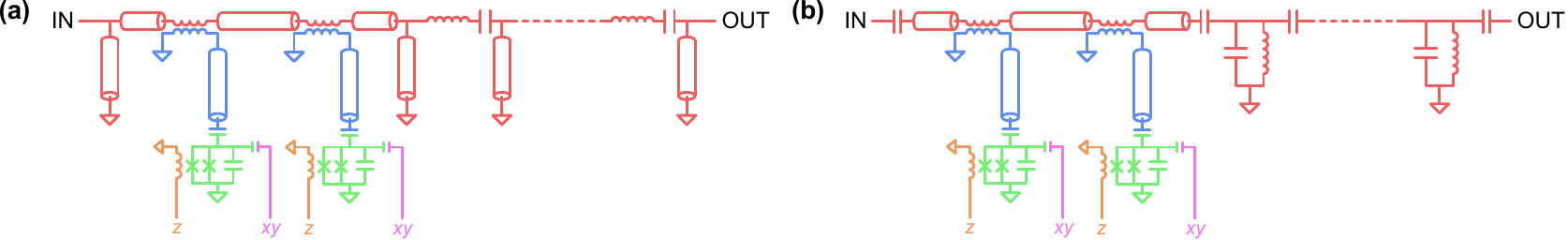}
    \caption{Alternative circuits for asymmetric filters with two readout resonators coupled to the first stage. (a) Using series $LC$ resonators on all the stages except the first stage and shorted-to-ground transmission line elements as the impedance inverters. (b) Using parallel $LC$ resonators on all the stages except the first stage and capacitors as the admittance inverters.}\label{figS9}
\end{figure*}
In the main text, we discussed an experimental realization of the filter using only sections of transmission lines, which is easy to design and implement. However, for an $N$th order filter, the layout will contain $N$ $\lambda/2$ transmission line resonators. A typical $\lambda/2$ transmission line resonator patterned on a sapphire wafer is around $6~\mathrm{mm}$ long. For higher order filters, the layout will take too much space. Here we present some alternative realizations with smaller footprints in Fig.~\ref{figS9}. For the stage coupled with the readout resonators (first stage for asymmetric filters and central stage for symmetric filters), we still use $\lambda/2$ transmission line resonators to couple more readout resonators. For other stages, we use series $LC$ (Fig.~\ref{figS9}(a)) or parallel $LC$ resonators (Fig.~\ref{figS9}(b)).
\bibliography{bpf_ref}